\documentclass[useAMS,usenatbib]{mn2e}

\usepackage{graphicx}
\usepackage{remreset} 
\usepackage{multicol}
\usepackage{amsmath}
\usepackage{float}

\usepackage{fixltx2e}
\usepackage{url}

\title[The LF and Local Hole]{The galaxy luminosity function and the Local Hole}
\author[J.~R. Whitbourn, T. Shanks]{J.R. Whitbourn$^{1}$\thanks{E-mail:
(JRW) jrwhitbourn@gmail.com} and T. Shanks$^{1}$\thanks{E-mail: (TS)
tom.shanks@durham.ac.uk} \\
$^{1}$Extragalactic Astronomy Group, Department of Physics, Durham
University, South Road, Durham, DH1 3LE, UK}

\begin{document}
\date{Accepted 2016 March 4. Received February 9; in original form 2015 June 22}

\pagerange{\pageref{firstpage}--\pageref{lastpage}} \pubyear{}
\maketitle

\label{firstpage}


\begin{abstract}

\citet{whitbourn_2013} have reported evidence for a local void
underdense by $\approx15$\% extending to 150-300h$^{-1}$Mpc around our
position in the Southern Galactic Cap (SGC). Assuming a local luminosity
function they modelled \textit{K}- and \textit{r}-limited number counts and
redshift distributions in the 6dFGS/2MASS and SDSS redshift surveys and derived
normalised n(z) ratios relative to  the standard homogeneous
cosmological  model. Here we test further these results using maximum
likelihood techniques that solve for the galaxy density distributions
and the galaxy luminosity function simultaneously. We confirm the
results from the previous analysis in terms of the number density
distributions, indicating that our detection of the `Local Hole' in the
SGC is robust to the assumption  of either our previous, or newly
estimated,  luminosity functions. However, there are discrepancies with
previously published \textit{K} and \textit{r} band luminosity functions. In
particular the \textit{r}-band luminosity function has a steeper faint end slope
than the $r_{0.1}$ results of \citet{blanton_2003} but is consistent with the
$r_{0.1}$ results of \citet{montero_2009,loveday_2012}.

\end{abstract}



\section{Introduction}
\label{sec:intro_locallf}

Our local galaxy clustering environment has recently assumed even greater
importance with the discovery that the SNIa Hubble diagram can be fitted by a
Universe with an accelerating expansion rate
\citep{schmidt_1998,perlmutter_1999}. Given the finely tuned nature of
the vacuum energy that is implied by cosmological explanations of the form of
the Hubble diagram, \citep{carroll_2001}, there is clear motivation to
look for other explanations for this observation. This has led to a variety of
activity investigating whether the local expansion rate is faster than at larger
distances due to the presence of a Local Hole or Void. Indeed, there have been
claims of a local underdensity manifesting as a local rise in SNIa based
measurements of $H_{0}$ \citep{zehavi_1998,jha_2007}.
Although some authors attribute these results to systematics associated with
dust \citep{conley_2007} these results are consistent with other work where bulk
flows out to $z<0.06$ are found using SNIa
\citep{feindt_2013,colin_2011,wojtak_2013} and the tension between local and CMB
determinations of $H_{0}$ (\citealt{planck_parameters_2013} - XVI). Some of the
work in this regard has even focused on non-Copernican models with the Local
Group positioned at the centre of a large void
\citep{clarkson_2010,schwarz_2010_new,krasinski_2013}. 
Here we are investigating a simpler scenario where the Local Group is at the
edge of an underdense region that covers much of the Southern Galactic Cap
(SGC). Evidence for such a possibility has been presented by
\citet{shanks_1990}, \citet{zucca_1997}, \citet{metcalfe_2001,metcalfe_2006},
\citet{busswell_2004} and
\citet{frith_2003,frith_2005a,frith_2005b,frith_2006b}. 

Whitbourn \& Shanks (2014, the compansion study to this paper, which we will
refer to hereafter as Paper I) have also recently presented evidence for a
local void with an $\approx 15$\% under-density out to
$\approx150-300$h$^{-1}$Mpc. These authors used 6dFGS/2MASS and SDSS redshift
surveys to probe the local region by modelling the $n(z)$ distributions from
three large regions of sky covered by these surveys. They also used the
$\overline{z}(m)$  technique of \citet{soneira_1979} to make a Hubble diagram
based on the redshift survey galaxies and showed that the data preferred a model
that showed coherent bulk motion out to 150h$^{-1}$Mpc compared to a model where
the galaxy motions recovered the CMB dipole within the survey region.

More recently \citet{keenan_2010,keenan_2012}; and \citet{keenan_2013}, have
compared galaxy counts and luminosity density at high and low redshift and
reported evidence for a 300Mpc void with a 50\% underdensity. Alternative probes
than \textit{K}-band galaxy surveys have also been used to study this
hypothesis. In particular \citet{boehringer_2014} used the X-ray selected REFLEX
II cluster survey. These authors find evidence for significant underdensities
with conclusions broadly similar to those of Paper I.


In Paper I we traced the local $n(z)$ using techniques that assumed the form of
the luminosity function (LF) from previous work. The assumed form was also
inferred in the $r$ and $K$ bands from original observations of LF's as a
function of galaxy morphology/$B-V$ colour in the $B$ band. Here we return
to the issue of the Local Hole now using maximum likelihood (ML) methods
\citep{choloniewski_1987, efstathiou_1988,cole_2011} that solve for the galaxy
density run with redshift, $\phi(z)$, simultaneously with solving
non-parametrically for the luminosity function. The only parameters needed are
simple forms for the \textit{K}-correction and evolution $K+E$ terms.

In particular, we begin by describing the techniques used in estimating the
galaxy LF and the underlying density fields. We first report the $V/V_{max}$
results for the K-band and relate these results to the number count slopes
reported in Paper I. We then show the \textit{K}-band LFs and compare to
the \citet{metcalfe_2001,metcalfe_2006} LF assumed in determining the density
profiles presented in Paper I. We proceed by presenting the density profiles
estimated in conjunction with the LF's using ML methods. We also include similar
results for the r band SDSS sample.

Throughout this paper we use a flat $\Lambda$ cold dark matter
($\Omega_{\Lambda,0}=0.7, \Omega_{m,0}=0.3$) cosmology with Hubble constant
$H=100$h kms$^{-1}$ Mpc$^{-1}$ with $h=0.7$.

\section{Techniques}
\label{subsec:techniques_lf}

We now briefly describe the methods of estimating the galaxy LF used in this
paper. Unless otherwise stated we have estimated non-parametric LFs using
binsizes of d$M=0.5$ and d$\mu=0.25$.

\subsection{Non-parametric luminosity function estimation}

\subsubsection{$V_{max}$ luminosity function}
\label{subsubsec:vmax_lf_esti}

We have used the standard $1/V_{max}$ estimator \citep{kafka_1967,schmidt_1968}.
This method assumes a homogeneous model and estimates the LF as

\begin{equation}
 \phi(M) = \sum_{i}^{N} \frac{1}{V_{i,max}} W(M_{i}-M),
 \label{eq:vmax_fundamental_eq}
\end{equation}

\noindent where $V_{max}$ is the comoving volume associated with the maximum redshift this
galaxy could be observed and $W(M_{i}-M)$ is a window function describing the
binning d$M$ assumed for the LF, i.e.

\begin{equation}
W(M_{i}-M) = %
\begin{cases}
1 & \text{if } -\textrm{d}M/2 \leq (M_{i}-M) \leq \textrm{d}M/2 \\
0 & \text{if else}
\end{cases}
\label{eq:windowfct_vmax_lf}
\end{equation}

\noindent One advantage of this method is the relative ease with which it can be
extended to allow weighting of galaxies. This can be achieved by replacing
the unity argument of the window function for galaxies in the absolute magnitude
bin d$M$ by a weighting factor \citep{ilbert_2005}. We have used the magnitude
dependent completeness factor described in appendix II of Paper I, i.e.
$1/f(m_{i})$ where $f(m_{i})$ is the spectroscopic success function described
in Paper I. We can account for a bright magnitude limit by replacing $V_{max}$
by $V_{max}-V_{min}$ in the denominator of equation
(\ref{eq:vmax_fundamental_eq}) since this is now the volume over which the
survey is complete at this absolute magnitude. 

Whilst this estimator of the LF is minimum-variance and ML it is also biased
as it assumes homogeneity and will therefore be affected by LSS
\citep{felten_1976}. Importantly, other LF estimators are unaffected by LSS
variations hence the difference between this LF estimator and
the others is therefore reflective of the presence of LSS. Although this method
offers an estimate of the global normalisation of the LF, no
estimate of the density run $\phi(z)$ is available from this binned LF
estimator.

\subsubsection{NPML: Choloniewski-Peebles luminosity function}

An alternative approach is a non-parametric maximum likelihood (NPML)
method due to PJE Peebles (private communication) and \citet{choloniewski_1986}.
The NPML method assumes separable densities $\rho_{i}$ and LF
$\phi_{j}$ with Poisson distribution in the brightness-distance modulus plane
$(M,\mu)$. The probability for $n_{i,j}$ galaxies to occupy the $i,j$th
brightness-distance modulus bin is

\begin{equation}
 p(N=n_{ij})=\frac{\exp^{-\rho_i\phi_j}(\rho_i\phi_j)^{n_{ij}}}{n_{ij}!}.
\end{equation}

\noindent Differentiating the log likelihood formed from these probabilities
gives estimates that can be solved iteratively \citep{takeuchi_2000}

\begin{align}
 \rho_{i} &= \sum_{j} n_{ij}/\sum_{j} \phi_{j}  \\
 \phi_{j} &= \sum_{i} n_{ij}/\sum_{i} \rho_{i}
\end{align}

\noindent On the basis that the cross terms are zero the Fisher matrix errors are
simply 

\begin{align}
 \sigma_{\rho_{i}} &= \frac {\rho_{i}}{(\sum_{j} n_{ij})^{1/2}}  \\
 \sigma_{\phi_{j}} &= \frac {\phi_{j}}{(\sum_{i} n_{ij})^{1/2}}
\end{align}
 
\noindent This is an ML method which is independent of inhomogeneity
\citep{choloniewski_1986}. Furthermore, it also offers an estimate of the global
normalisation of the LF and the density run $\phi(z)$. However
the method's accuracy is dependent on galaxies being Poisson distributed across
the brightness and distance modulus binning. The validity of this assumption is
improved by smaller bin sizes but at the expense of possible bias (which
increases with smaller bin sizes; \citealt{choloniewski_1986}). We have used
d$M=0.5$, d$\mu=0.25$ for the \textit{K} band and d$M=0.2$, d$\mu=0.2$ for the
\textit{r} band.

\subsubsection{$C^{-}$ luminosity function}

We have also used the $C^-$ method of \citet{Lynden_1971} as updated by
\citet{choloniewski_1987}. Here the distribution of galaxies in the
$(M,\mu)$ plane is used to infer a binned non-parametric LF.

For a sample sorted from brightest to faintest we construct the $C^{-}$
statistic as \citep{Lynden_1971} follows,

\begin{equation}
 C^{-} = \sum_{i}^{M<M_{i}} \ \sum_{m_{b}-M_{i} \leq \mu_{j}}^{\mu_{j} \\ \leq
m_{f}-M_{i}} w_{ij},
\end{equation}

\noindent where $w_{ij}$ is the weight of each galaxy which can be used to
account for incompleteness \citep{ilbert_2005}. We have again accounted for
incompleteness as $w_{ij}=1/f(m_{ij})$ where $f(m_{ij})$ is the spectroscopic
success function described in Paper I. The summation is defined by the ranges
associated with the faint ($m_{f}$) and bright ($m_{b}$) magnitude limits.

These $C^{-}$ coefficients can then be related to the cumulative LF
$\Phi(M)$ through a recursion relation. This method for estimating the
cumulative LF was modified by \citet{choloniewski_1987} who extended it to
enable an estimate of the underlying LF with a global normalisation and a
density profile. It is this version of the estimator that we use in this study.
Further discussion of the method can found in \citet{choloniewski_1987},
\citet{willmer_1997} and \citet{takeuchi_2000}.

\subsubsection{Joint SWML method}

The Efstathiou-Ellis-Peterson (EEP) estimator is a ML estimate which maximises
the probability of selection \citep{efstathiou_1988}:

\begin{equation}
 p=\frac{\phi(M)}{\int_{M_{b}(z)}^{M_{f}(z)} \phi(M) dM}.
 \label{eq:jswml_probfunc}
\end{equation}

\noindent Here the bounds on the integral are defined by the selection criteria of the
survey. We can therefore calculate the likelihood, $\mathcal{L} = \prod_{i}^{N}
p_{i}$, over binned values of the LF to find the ML estimator.
This binning of the LF requires the use of step-functions in describing the
ML solution. This step-wise approach has led to the estimator
being described as the Step-Wise ML method (SWML). This method has been updated
by \citet{cole_2011} to jointly estimate the global normalisation, density
profiles and the LF (JSWML). It is this JSWML version of the non-parametric ML
method that is used in this study.

Our implementation of this method is based on a modified version of the
\texttt{JSWML} code provided by \citet{cole_2011}\footnote{These modifications
resolved issues with the absolute magnitude bin centres and the number of
redshift bins.}. We have used the default settings except for implementing the
K+E corrections described in Sec. \ref{subsec:kcorr_evol_lf}, the specific
cosmological parameters used in this paper and the absolute magnitude range
required.

\subsection{Parametric Luminosity Function estimation}

The estimation of LFs can be analytically simplified by assuming a parametric
form. This is typically achieved for galaxies by using a \citet{schechter_1976}
fit.

\subsubsection{STY luminosity function}

The Sandage-Tammann-Yahil (STY) method is akin to the EEP and JSWML methods in
that it is a ML estimator \citep{sandage_1979}. Here though we calculate the
likelihood, $\mathcal{L} = \prod_{i}^{N} p_{i}$, over a plane of possible values
of the Schechter parametrisations $(\alpha,M_{*})$ to find the ML estimates of
the LF parameters.

We have evaluated $\mathcal{L}(\alpha,M_{*})$ over $\alpha \in [0.8,1.6]$ and
have adapted the range for $M_{*}$ for each sample on the basis of the
estimated covariance matrix. In both cases we have used binsizes of
0.01. Incompleteness effects can be accounted for by weighting each probability
as $p_{i}^{w_{i}}$ where $w_{i}$ is the inverse of the spectroscopic
success function $f(m)$ described in Paper I. For a fuller discussion and a
full expansion of the log likelihood - see \cite{zucca_1994,zucca_1997}. We
note that this method estimates Schechter LF parameters $(\alpha,M_{*})$ but
does not provide any estimate of the global normalisation $\phi_{*}$ or the
density profile. It should also be noted that the accuracy of the STY LF
estimates is dependent on the validity/accuracy of the assumed parametric form.

\subsection{Luminosity function and Density Profile normalisation}

\subsubsection{Luminosity function normalisation}

An LF normalisation is related to the spatial number density as,

\begin{equation}
 n = \int_{-\infty}^{\infty} \phi(M) dM.
 \label{eq:numden_norm_phi}
\end{equation}

\noindent A minimum variance estimator ($n_{3}$) was found by
\citet{davis_1982} and has been commonly used. However, it is not an unbiased
estimator for inhomogeneous samples as the number density is also
present in the galaxy weighting. Although this effect is expected to be small
\citep{willmer_1997,keenan_2012}, we have decided to use an unbiased estimator
of the number density \citep{davis_1982}:

\begin{equation}
 n_{1} = \frac{\int_{0}^{z_{f}} (N(z)/s(z)) dz}{\int_{0}^{z_{f}} dV}.
 \label{eq:n1define}
\end{equation}

\noindent Here $s(z)$ is the galaxy selection function and $N(z)$ is the
redshift distribution of galaxies. The disadvantage of this estimator is the
instability associated with its heavier weighting of higher redshift objects
where the selection function is more uncertain. Various methods such as using
medians etc have been proposed for improving the robustness of these estimators
\citep{deLapparent_1989}. We consider a high redshift cut-off of $z_{f}=0.04$
and $z_{f}=0.075$ in the estimation of $n_{1}$ for the \textit{K} and
\textit{r}-band respectively (i.e. approximately the maximum of the respective
redshift distributions). The resulting unbiased estimator $n_{1}$ can then be
used to normalise $\phi(M)$ following equation (\ref{eq:numden_norm_phi}).

\subsubsection{Density Profile normalisation}

We have considered a variety of methods for normalising the density profiles.
In \citet{willmer_1997} it was shown that a number count type estimator is
relatively unbiased as compared to other ML density estimates. These
ML estimates showed an $\approx20\%$ bias towards underestimating density. The
results presented in this study are therefore based on a number count
normalisation derived for each respective LF estimate. The number count
normalisation has been made by estimating the change in $\phi_{*}$ required to
fit the number counts (as per the method in \citealt{efstathiou_1988}) and
scaling
the density profiles accordingly. 

We have ensured that these number count based results are consistent with the
$n_{1}$ estimator used in normalising the LF's by considering a number density
profile estimator derived from the $n_{1}$ estimator, i.e. the ratio of
the expected number of galaxies in a redshift shell of thickness $dz$ and the
volume of the redshift shell

\begin{equation}
 \frac{n(z)}{n_{1}} = \frac{N(z)/s(z)}{\frac{dV}{dz}dz}
 \label{eq:n1densityprof}
\end{equation}

\noindent The results obtained using the unbiased $n_{1}$ estimator are in
agreement with those shown, but with larger uncertainties. For further detail
of the techniques we use in estimating the LF and its normalisation - see
\citet{johnston_2011}.

\section{Data \& Modelling}
\label{sec:data_lf}

The imaging and redshift surveys used here are the same as those used in Paper
I, namely 2MASS \citep{jarrett_2003} and SDSS \citep{york_2000} for
near-infrared (NIR) and optical imaging and 6dFGS \citep{jones_2004} and SDSS
for \textit{K} and \textit{r} limited galaxy redshift surveys. We again adopt
the Vega photometric system and use the Local group rest frame whilst adopting
the transformations outlined in Paper I. We also reprise the magnitude
estimators used in Paper I, i.e. a scale error corrected form of the `k\_m\_ext'
magnitude for 2MASS objects and the `cmodel' magnitude for SDSS objects - see
\citet{whitbourn_2013} for further discussion. To minimise the effects of
incompleteness for the r-band sample we have employed the more conservative
magnitude selection $10<r<17$ than was used in Paper I. We have used an
expanded $7.5<K<12.5$ selection criteria for the K band. We now use a $7.5<K$
bright limit rather than the $10<K$ used in Paper I in order to maximise sample
completeness whilst avoiding the range affected by 2MASS deblending issues.

We have used a faint absolute magnitude limit of $M- 5\log h
\leq -18$ and $M - 5\log h \leq -15$ for the \textit{K} and \textit{r} band,
respectively. We have ensured the accuracy of our modelling procedures by
validating with respect to simulated data.

Within these surveys we again use the same large target fields as used in Paper
I (see fig. 1 and table 3 of Paper I). These regions are chosen so as to be
relatively similar in their dimensions, whilst being as large as their
constituent surveys' geometry allows a coherent field to be. The largest fields
possible were preferred since these should minimise cosmic variance (each
represents $\approx 1/15$ of the sky). These fields were also selected to
represent regions of interest such as the CMB heliocentric dipole pointing and
the Great Attractor whilst avoiding the galactic plane. 

We will use galactic coordinates to define the fields as being northern or
southern, and use the different surveys to further distinguish the two galactic
northern fields i.e.: SDSS-NGC (Northern Galactic Cap), 6dFGS-NGC and 6dFGS-SGC.

\subsection{K-corrections and Evolution}
\label{subsec:kcorr_evol_lf}

We have followed Paper I in assuming simple representations of the
\citet{bruzual_2003} K-correction plus evolution models as used by Metcalfe et
al to fit galaxy counts and colours to much higher redshifts than those
discussed here.

For simplicity we have used a simple representation for the $\tau=2.5$Gyr, $x=3$
and $\tau=9$Gyr, $x=1.35$  K+E corrections for early-type and late-type galaxies
in the K band. For both types there is little difference here between the K- and
K+E corrections out to $z\approx0.3$.

In the \textit{r}-band there is a bigger difference between the K and the K+E
corrections. We therefore use the average \textit{r}-band K+E-correction for
early-type and spirals assuming the same \citet{bruzual_2003} models as for $K$
above. Although there is a slight approximation here involved in taking an
average K+E/K-correction at the $z=0.1$ limit of the range of interest the
difference is only 0.05-0.06mags.

\subsection{Error Calculation}
\label{subsec:error_calc_lf}

To estimate random errors we use $10^\circ\times10^\circ$ subfields to
calculate jack-knife errors as used in Paper I. We found the number of subfields
only weakly effect error estimates. The only exception to this is our use of Fisher
matrix errors in the case of the NPML LF estimator.

\section{Luminosity Functions}
\label{sec:lfcalc}

\subsection{$V/V_{max}$ Histograms}

Before studying the LF estimates we first probe the $V/V_{max}$ statistic. This
is of particular interest because it is closely related to the $1/V_{max}$ LF
estimator but is also dependent on the homogeneity of the sample - see Sec.
\ref{subsubsec:vmax_lf_esti}. The $V/V_{max}$ statistic has been calculated
using the $7.5<K<12.5$ selection criteria, incompleteness correction and the
K+E prescription outlined in Sec. \ref{subsec:kcorr_evol_lf}. The homogeneous
expectation is therefore that the samples are uniformly distributed over the
volume probed and hence the mean $V/V_{max}=0.5$. 

\begin{figure}
   \centering
\includegraphics[width=0.5\textwidth]{%
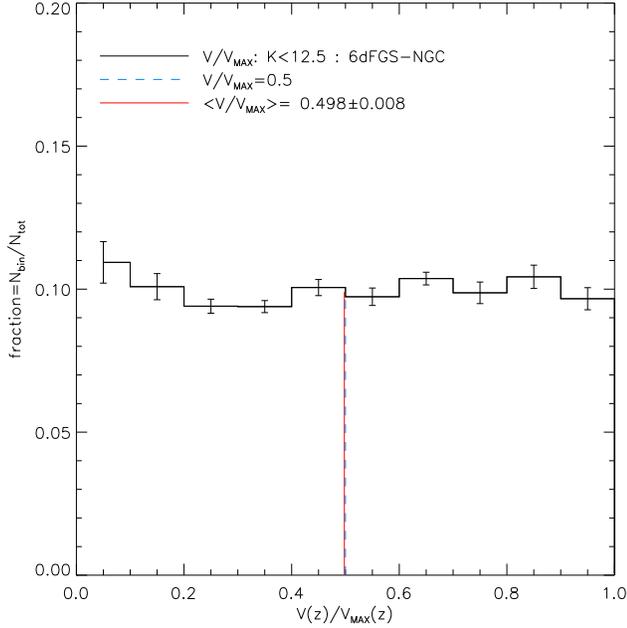}
    \caption[]{A histogram of \textit{K} band galaxy $V/V_{max}$ with
$7.5<K<12.5$ for the 6dFGS-NGC data with corresponding jackknife error. The
(blue, dashed) line shows the homogeneous expectation that $\langle V/V_{max}
\rangle=0.5$. Also shown is the mean $\langle V/V_{max} \rangle$ for 6dFGS-NGC
data (red, solid line).}
    \label{fig:vvmax_sixdf_north_K}
\end{figure}

\begin{figure}
   \begin{minipage}{0.5\textwidth}
   \centering
\includegraphics[width=1.0\textwidth]{%
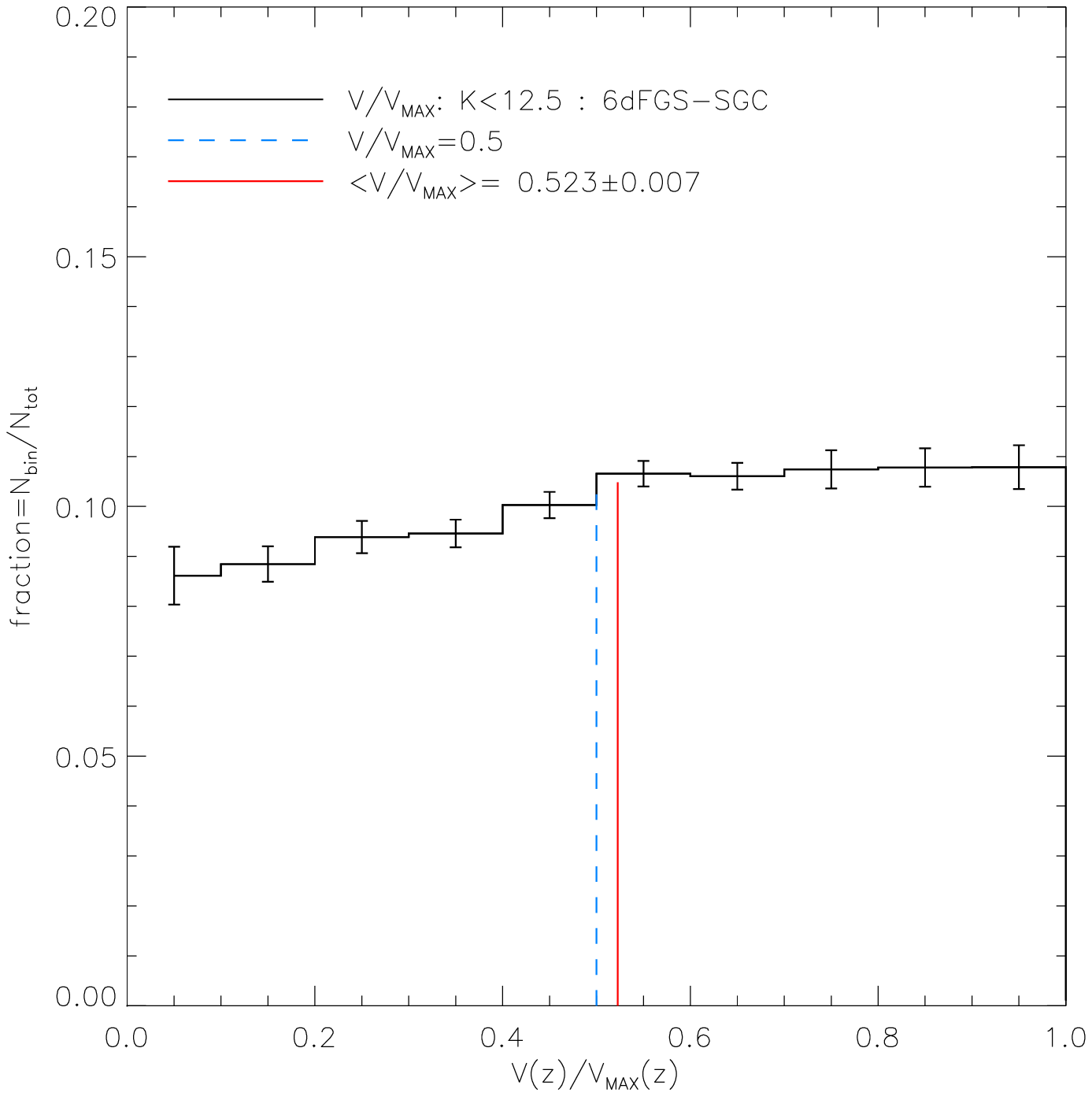}
    \caption[]{A histogram of \textit{K} band galaxy $V/V_{max}$ with
$7.5<K<12.5$ for the 6dFGS-SGC data with corresponding jackknife error. The
(blue, dashed) line shows the homogeneous, expectation that $\langle V/V_{max}
\rangle =0.5$. Also shown is the mean $\langle V/V_{max} \rangle$ for 6dFGS-SGC
data (red, solid line).}
    \label{fig:vvmax_sixdf_south_K}
    \end{minipage}
    \hspace{1pc}%
    \hfill
    \begin{minipage}{0.5\textwidth}
    \centering
\includegraphics[width=1.0\textwidth]{%
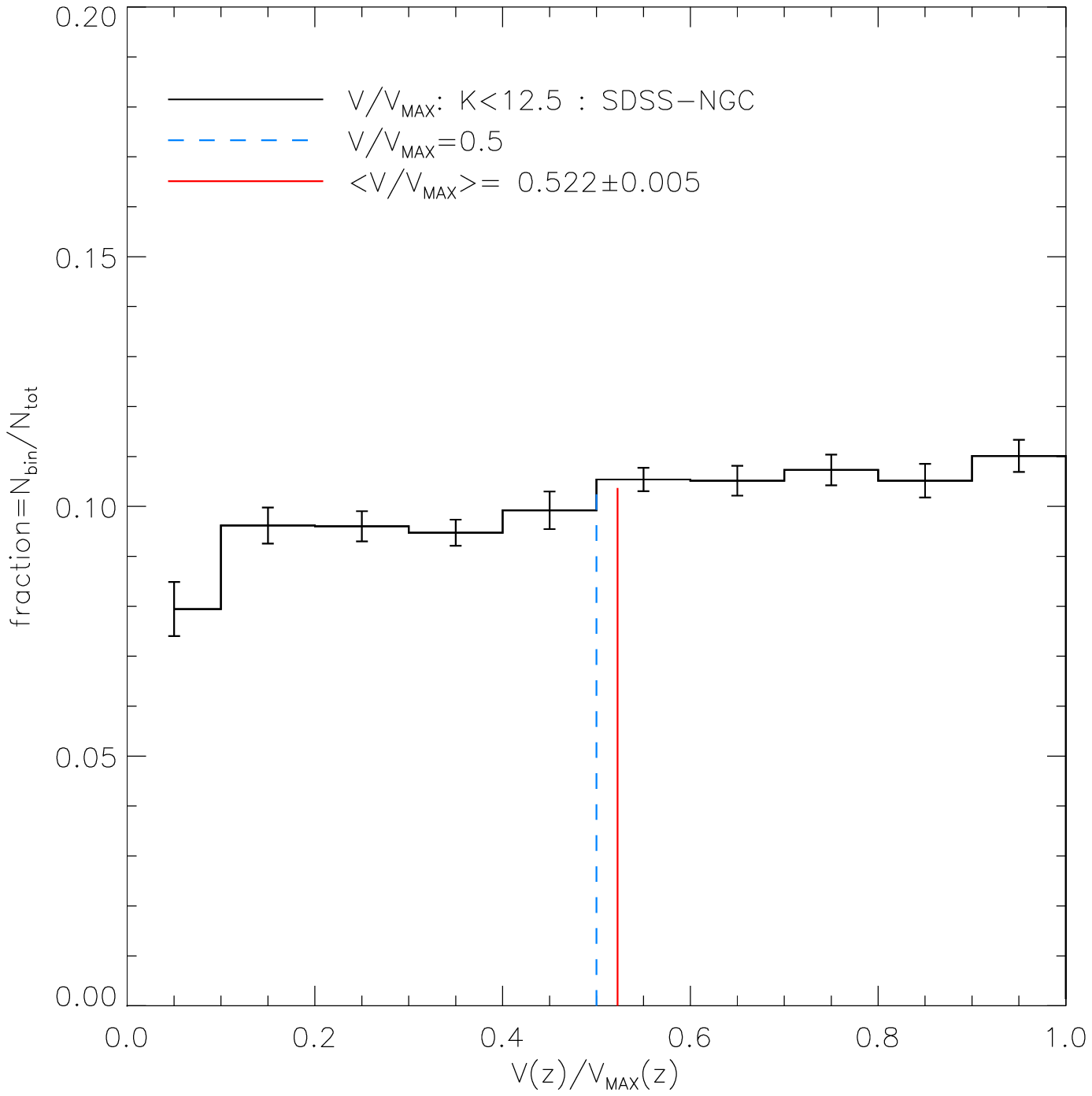}
    \caption[]{A histogram of \textit{K} band galaxy $V/V_{max}$ with
$7.5<K<12.5$ for the SDSS-NGC data with corresponding jackknife error. The
(blue, dashed) line shows the homogeneous expectation that $\langle V/V_{max}
\rangle=0.5$. Also shown is the mean $\langle V/V_{max} \rangle$ for SDSS-NGC
data (red, solid line).}
    \label{fig:vvmax_sdss_north_K}
\end{minipage}
\end{figure}

We now show in Figs \ref{fig:vvmax_sixdf_north_K}-\ref{fig:vvmax_sdss_north_K}
a histogram of this statistic for the K band data over our three target regions
with a binning of d$(V/V_{max})=0.1$. We find for the 6dFGS-NGC, 6dFGS-SGC and
SDSS-NGC regions mean values of $V/V_{max}$ of $(0.498\pm0.008)$,
$(0.523\pm0.007)$ and $(0.522\pm0.005)$ respectively. We conclude that in the
6dFGS-SGC and SDSS-NGC regions the data is not consistent with a uniform
distribution and is in fact increasing with $V/V_{max}$. Given that
incompleteness effects have been included in the calculation of $V_{max}$ the
significant excess above the homogeneous prediction in the 6dFGS-SGC and
SDSS-NGC regions indicates that these samples are being preferentially
distributed at higher redshifts. We therefore conclude that there
is significant evidence for an inhomogeneity, and in particular a local
underdensity, on the basis of the $V/V_{max}$ statistic alone. 

We also note that the sloping of the $V/V_{max}$ statistic is closely related to
the rising number counts of these samples as was observed in Paper I. Indeed it
is the 6dFGS-SGC region which has the most pronounced sloping in $V/V_{max}$ and
was the most underdense in Paper I. Clearly however determining the density
profile and its run with redshift requires solving for the density profile. But
first we now investigate whether the LF of these samples is consistent with
those of the \citet{metcalfe_2001,metcalfe_2006} LF assumed in determining the
density profiles presented in Paper I.


\subsection{\textit{K} band LF estimates}
\label{subsec:Kband_lf_Estimates}

\begin{figure}
   \centering
\includegraphics[width=0.5\textwidth]{%
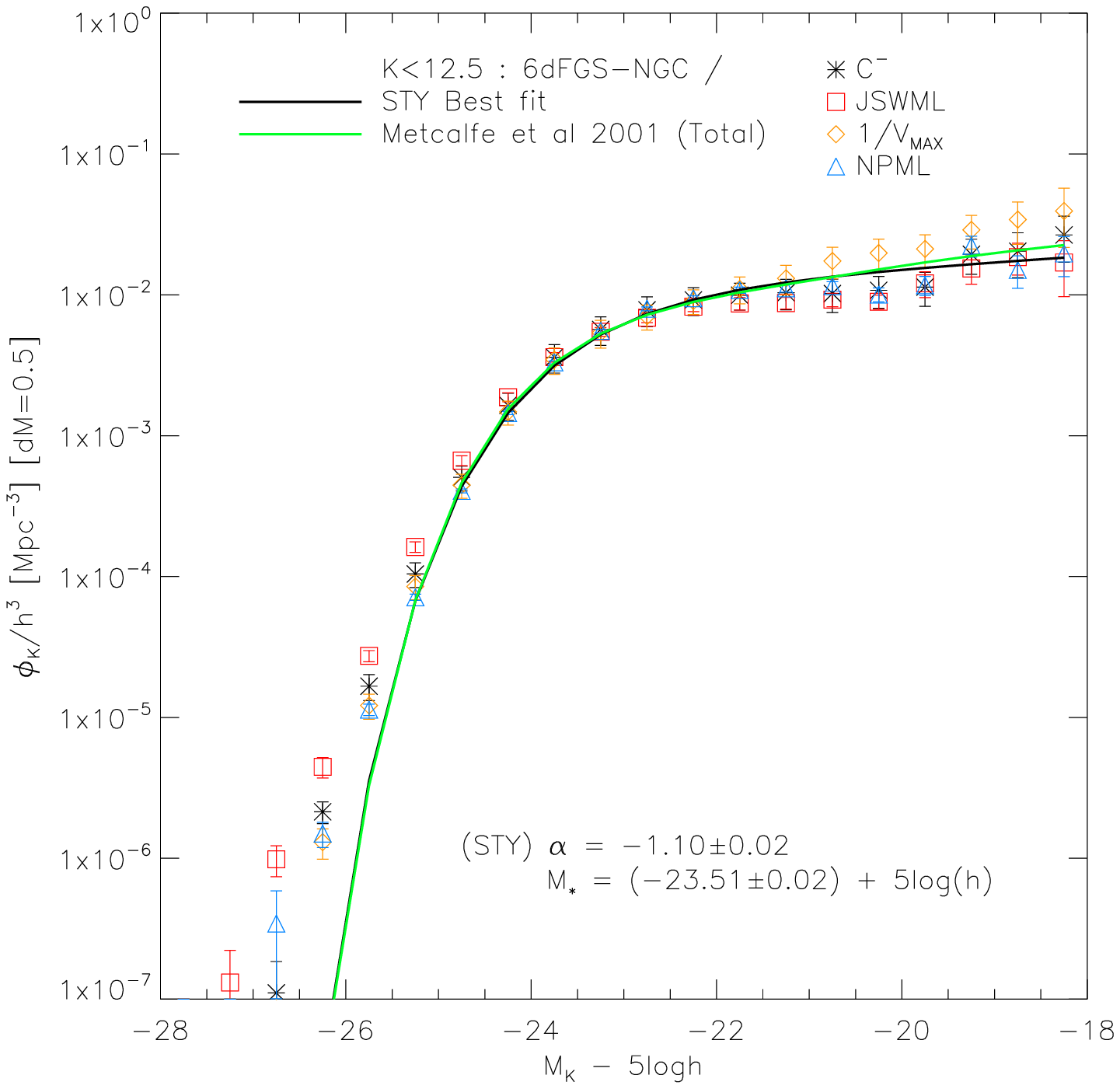}
    \caption[]{\textit{K} band galaxy LF estimates with $7.5<K<12.5$ and
d$M=0.5$ for the 6dFGS-NGC data. All of the estimates have been normalised using
the $n_{1}$ number density estimator for their respective LF.}
    \label{fig:lf_sixdf_north_K}
\end{figure}
\begin{figure}
    \centering
\includegraphics[width=0.5\textwidth]{%
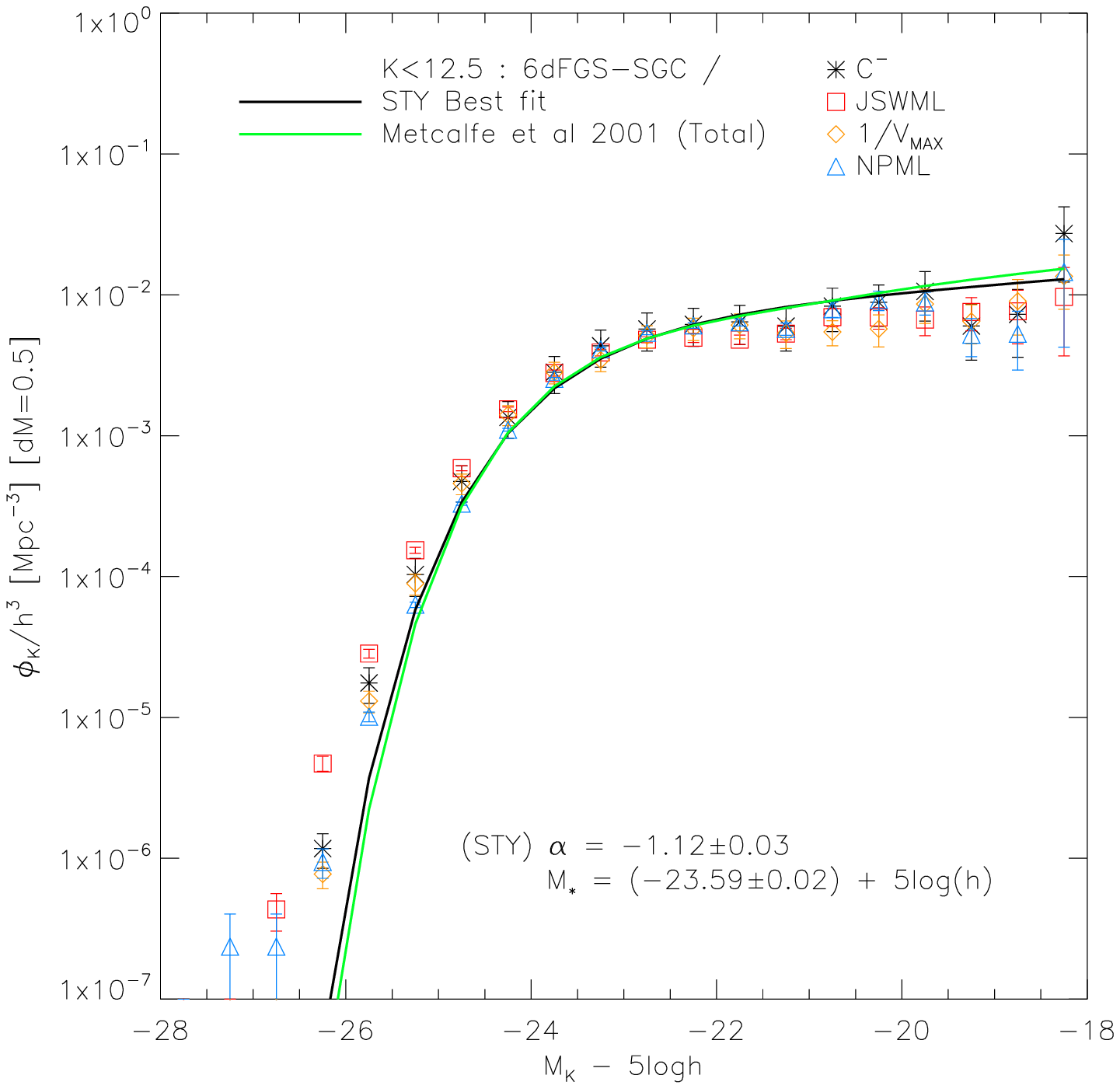}
    \caption[]{\textit{K} band galaxy LF estimates with $7.5<K<12.5$ and
d$M=0.5$ for the 6dFGS-SGC data. All of the estimates have been normalised using
the $n_{1}$ number density estimator for their respective LF.}
    \label{fig:lf_sixdf_south_K}
\end{figure}
\begin{figure}
   \centering
\includegraphics[width=0.5\textwidth]{%
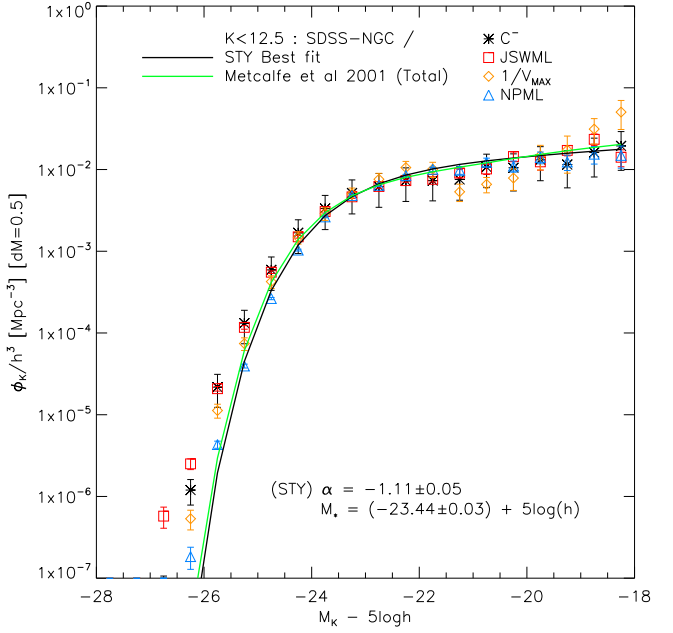}
    \caption[]{\textit{K} band galaxy LF estimates with $7.5<K<12.5$ and
d$M=0.5$ for the SDSS-NGC data. All of the estimates have been normalised using
the $n_{1}$ number density estimator for their respective LF.}
    \label{fig:lf_sdss_north_K}
\end{figure}

We next use the LF estimators described in Sec. \ref{subsec:techniques_lf} to
estimate the \textit{K} band LF in our three regions. In Figs
\ref{fig:lf_sixdf_north_K}-\ref{fig:lf_sdss_north_K} we show these estimates
for the 6dFGS-NGC, 6dFGS-SGC and SDSS-NGC regions, respectively. These LF's
have been normalised using their respective estimate of $n_{1}$.

Treating these fields in turn, we begin with the 6dFGS-NGC field shown in Fig.
\ref{fig:lf_sixdf_north_K}. We first note that the LF estimators
are in agreement with the \citet{metcalfe_2001,metcalfe_2006} type dependent
Schechter LF (green solid line) until the very bright end ($<M_{*}-2$). The
Metcalfe et al LF is an optical LF which is translated into the NIR using an
assumed mean colour so this agreement was not to be taken for granted. We
also note that the parametric Schechter function fits provided by the STY method
is in agreement with the other non-parametric LF estimates over much of the
range in absolute magnitude. However, for the very bright end ($<M_{*}-2$) the
STY result is an underestimate with respect to the nonparametric estimates. We
interpret this as suggesting that Schechter parametrisation of the LF is
accurate in the main but does not fully represent the abundance of very bright
objects such as brightest cluster galaxies - a conclusion also reached by other
authors \citep{jones_2006}.

For the 6dFGS-SGC region shown in Fig. \ref{fig:lf_sixdf_south_K} we see a
similar set of estimates to those obtained in the 6dFGS-NGC region up until 
the faint end ($<M_{*}+2$). The Metcalfe et al LF is a reasonable fit to all
the estimators over all but the brightest and faintest magnitudes. Indeed, this
time the parametric estimator, the STY method, agrees well with the
non-parametric estimates except for $<M_{*}-2$ and $<M_{*}+2$. We again assign
this relative excess of faint objects and deficit of bright objects relative to
the nonparametric fits as due to a limitation of this parametrisation. We also
note that the $V_{max}$ estimator is significantly different from the other
estimates. We attribute this to the inhomogeneity reported in this field in
Paper I. This is consistent with the elevated value of
$\langle V/V_{max} \rangle =(0.523\pm0.007)$ in this region as discussed in Sec.
\ref{sec:lfcalc}.

Now in Fig. \ref{fig:lf_sdss_north_K} we show the LF estimators for
SDSS-NGC region. We draw similar conclusions to the previous 6dFGS-SGC region.
In particular the $V/V_{max}$ LF estimate is again significantly different from
the other estimators, especially at the faint end. We again attribute this to a
significant inhomogeneity as indicated in this field by the elevated value of
$\langle V/V_{max} \rangle=(0.522\pm0.005)$ discussed in Sec. \ref{sec:lfcalc}. 
We note that the bright end excess relative to the Schechter parametric fits
is less pronounced in this field. We also observe that whilst the STY $\alpha$
estimates are mutually consistent between the fields the differences in the
$M_{*}$ estimates, although small, are significant. But in any case the
differences are relatively minor in the sense that the Metcalfe et al LF is a
good representation of all the LF estimators except at the bright end.

We therefore conclude that the Metcalfe et al LF is a adequate fit to the
majority of the \textit{K} band LF estimators in all three fields. We take
this as providing evidence supporting our assumption of the Metcalfe et al LF in
Paper I - and hence the density profiles we then estimated using number counts.
However, we can now continue this investigation beyond the LF and make use of
the density profiles and normalisation information provided by the various LF
estimates to study the homogeneity of our samples.

\section{K band normalised number density profiles}
\label{sec:lf_densprof_calc}

\begin{figure}
   \centering
\includegraphics[width=0.5\textwidth]{%
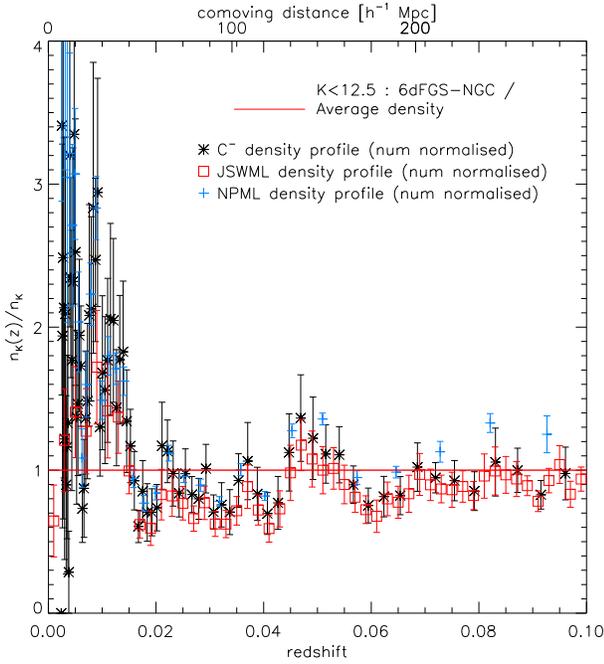}
    \caption[]{\textit{K} band galaxy number underdensity profile with
$7.5<K<12.5$ for the 6dFGS-NGC data normalised (for each LF) to the number
counts in the region. The red (solid) line represents the homogeneous
prediction for each LF.}
    \label{fig:nz_sixdf_north_K1}
	    \end{figure}
	    \begin{figure}
\includegraphics[width=0.5\textwidth]{%
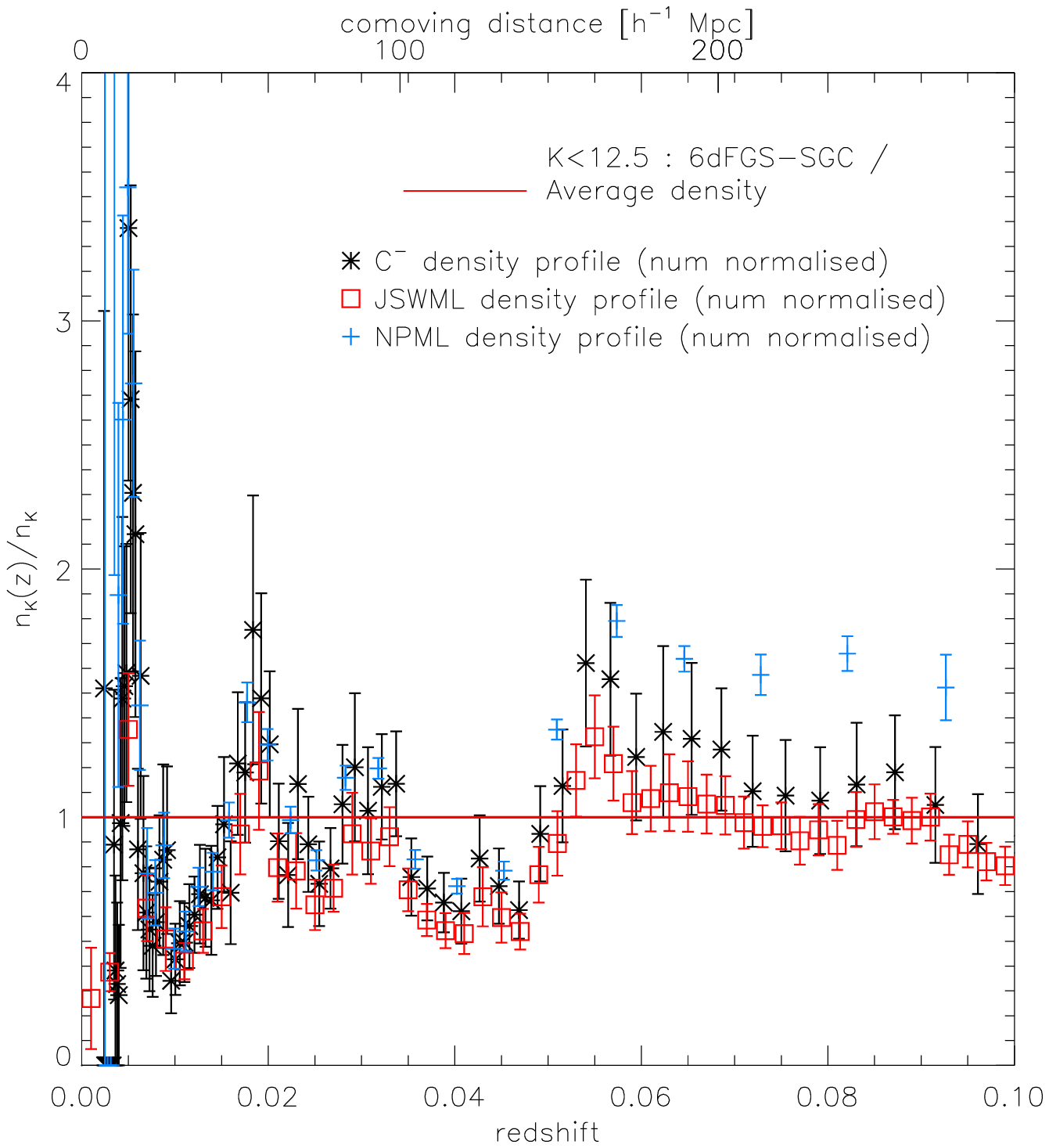}
    \caption[]{\textit{K} band galaxy number underdensity profile with
$7.5<K<12.5$ for the 6dFGS-SGC data normalised (for each LF) to the number
counts in the region. The red (solid) line represents the homogeneous
prediction for each LF.}
    \label{fig:nz_sixdf_south_K2}
\end{figure}
\begin{figure}
   \centering
\includegraphics[width=0.5\textwidth]{%
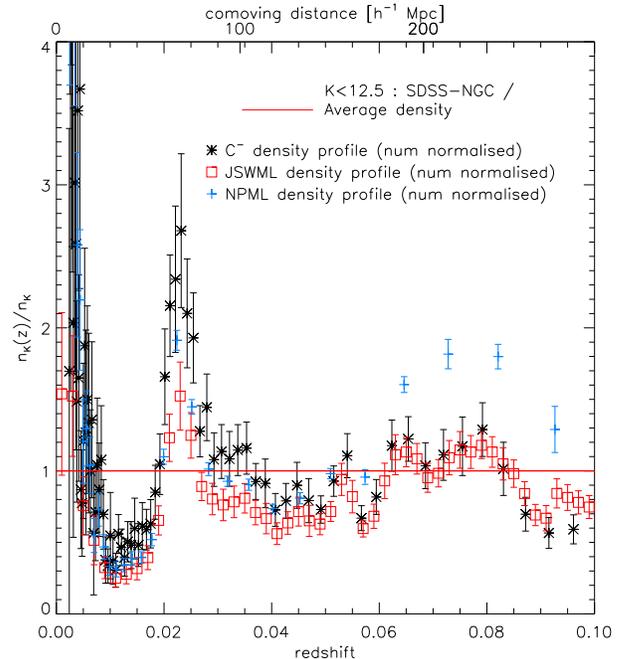}
    \caption[]{\textit{K} band galaxy number underdensity profile with
$7.5<K<12.5$ for the SDSS-NGC data normalised (for each LF) to the number counts
in the region. The red (solid) line represents the homogeneous prediction for
each LF.}
    \label{fig:nz_sdss_north_K3}
\end{figure}

As outlined in Sec. \ref{subsec:techniques_lf} we can use the NPML, JSWML and
$C^{-}$ methods to estimate the run of the number density profile. The
normalisations used are dependent on the estimator. The profiles presented here
have been normalised using their respective LF based number counts. Similar
results, at greater uncertainty, are found when using the $n_{1}$ unbiased
number density estimator (i.e. equation \ref{eq:n1densityprof}) calculated using
the corresponding LF. We now present in Figs
\ref{fig:nz_sixdf_north_K1}-\ref{fig:nz_sdss_north_K3} these number density
profiles for the 6dFGS-NGC, 6dFGS-SGC and SDSS-NGC regions, respectively. 

For the 6dFGS-NGC region (Fig. \ref{fig:nz_sixdf_north_K1}) we observe good
consistency between the various estimates of the number density run. We can also
see the inhomogeneity inferred from this field's $V/V_{max}$ statistic in that
locally ($z<0.05$) the profiles are typically underdense and then
transition to being reasonably homogeneous. We also observe significant LSS
clustering with significant fluctuations in the density profile at
$z\approx0.048$ which we attribute to the Shapley-8 supercluster. We also note
that this profile is similar to that presented in Paper I (fig. 3a) for the
6dFGS-NGC field where the Metcalfe et al LF was assumed. This is inline with the
agreement we noted in Sec. \ref{subsec:Kband_lf_Estimates} between the Metcalfe
et al. LF and the LF estimates we have made here.

In Fig. \ref{fig:nz_sixdf_south_K2} we show the number density profile for the
6dFGS-SGC field. We again see good agreement between the different estimators
of the density profile. However, at high redshift ($z>0.05$) the NPML estimate
is significantly higher than the JSWML and $C^{-}$ estimates. This is true
for all three \textit{K}-band fields and may therefore be indicative of a lack
of robustness at high redshift for the NPML estimate. All profile estimates are
particularly underdense at local redshift with the JSWML and $C^{-}$ becoming
homogeneous at deeper redshifts. We again note the agreement with the
Paper I (fig. 3b) number density profile for this field  which once more
reflects the validity of the Metcalfe et al LF for the sample in this field.

Finally, in Fig. \ref{fig:nz_sdss_north_K3} we show the number density profile
for the SDSS-NGC region. Here we see a similar pattern of agreement between the
different number density profiles. The JSWML, NPML and $C^{-}$ profiles are in
agreement at low redshift in showing an underdense profiles with significant LSS
(which we attribute to Coma at $z=0.023$). At deeper redshifts ($z>0.08$), the
density profiles show evidence of an extensive overdensity. We return to
investigate this issue using the deeper \textit{r} band data over the same
field. However, we also note that this substantially inhomogeneous profile is in
agreement with the density profile estimated in Paper I (fig. 3c).

We have evaluated the corresponding number under-density indicated by these
profiles as

\begin{equation}
  \frac{\int_{0}^{z_{1}} \frac{n(z)}{n_{K}} \frac{dV}{dz}.
dz}{\int_{0}^{z_{1}} dV}
 \label{eq:density_evaluate}
\end{equation}

The average of the profile number underdensities are reported in Table
\ref{tb:underdensity_table_K} for $z<0.05$, i.e. $\approx150$h$^{-1}$Mpc, and
$z<0.1$, i.e. $\approx300$h$^{-1}$Mpc in the 6dFGS-NGC, 6dFGS-SGC and SDSS-NGC
regions. Errors have been inferred using jackknife estimates. We noted earlier
the potential lack of robustness at high redshift for the NPML estimator. We
have therefore disregarded the NPML profiles in calculating the $z<0.1$ average
underdensities.

These results are broadly consistent with the underdensities reported in Paper
I (table 4) - aside from the 6dFGS-SGC estimate. In this case, both the
$z<0.05$ (now $0.76\pm0.05$, previously $0.60\pm0.05$) and $z<0.1$ results
(now $1.02\pm0.11$, previously $0.75\pm0.04$) are less underdense. Indeed the
$z<0.1$ result is now consistent with homogeneity, albeit with larger errors.
However, this difference is relatively minor in that the results presented here
based on number count normalised number density profiles are in agreement with
the 6dFGS-SGC number count underdensity reported in Paper I ($0.76\pm0.03$).

We therefore conclude that the density profiles show evidence for an LSS local
underdensity for $z<0.05$ which the SDSS-NGC field in particular suggests may
extend to deeper depths ($\approx300$h$^{-1}$Mpc). These conclusions are in
agreement with those presented in Paper I which reflects the agreement found
here with the Metcalfe et al LF used in that study.

\begin{table}
 \begin{center}
 \begin{tabular}{ccc}

\hline
Field & Sample limit & Under-density \\
\hline

6dFGS-NGC & $z<0.05$  & $0.95\pm0.11$ \\

6dFGS-SGC & $z<0.05$ & $0.76\pm0.05$ \\

SDSS-NGC & $z<0.05$ & $0.83\pm0.05$ \\

\hline

6dFGS-NGC & $z<0.1$  & $0.91\pm0.08$ \\

6dFGS-SGC & $z<0.1$ & $1.02\pm0.11$ \\

SDSS-NGC & $z<0.1$ & $0.89\pm0.06$ \\

\hline
\end{tabular}
\caption{A summary of the average underdensities derived using eqn.
(\ref{eq:density_evaluate}). The $z<0.05$ and $z<0.1$ entries use
$7.5<K<12.5$.
 }
\label{tb:underdensity_table_K}
 \end{center}
\end{table}

\section{\textbf{$\MakeLowercase{r} \leq 17$} LF and density profiles in the
SDSS-NGC region}
\label{sec:rband_sdss_work}

Using the SDSS survey it is possible to go to deeper survey limits than in
the \textit{K}-band. In particular, following Paper I, we use an \textit{r} band
limited sample in order to investigate the SDSS-NGC field. We have used
a more conservative r-band magnitude limit of $r \leq 17$ than the $r \leq 17.2$
limit used in Paper I in order to minimise any potential biasing/issues
associated with spectroscopic incompleteness. 

\begin{figure}
   \centering
\includegraphics[width=0.5\textwidth]{%
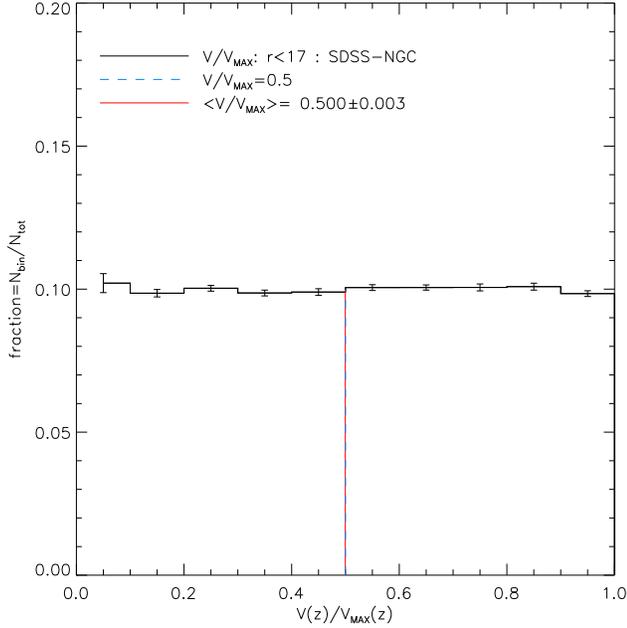}
    \caption[]{A histogram of \textit{r} band galaxy $V/V_{max}$ with
$10<r<17$ for the SDSS-NGC data with a corresponding jackknife error. The (blue,
dashed) line shows the homogeneous, complete expectation that $\langle V/V_{max}
\rangle=0.5$. Also shown is the mean $\langle V/V_{max} \rangle$ for SDSS-NGC
data (red, solid line).}
    \label{fig:vvmax_sdss_north_r}
%
\end{figure}

Now in Fig. \ref{fig:vvmax_sdss_north_r} we show the $V/V_{max}$ estimates. We
report a mean value of $(0.500\pm0.003)$. This value is consistent with the
homogeneous expectation and indeed the uniformity in this statistic shows little
evidence for an underdensity. This is significantly different from the
slope observed in Fig. \ref{fig:vvmax_sdss_north_K} where the \textit{K} band
SDSS-NGC $V/V_{max}$ estimates are shown. A similar and related situation was
encountered in Paper I when comparing the results \textit{r} and \textit{K} band
counts over the same field. Here it was concluded that the evidence for an
underdensity in the \textit{r} band was more ambiguous and more suggestive of
$150h^{-1}$ scale underdensity which was punctuated by the strong clustering
associated with the Coma supercluster. This is potentially consistent with the
broader smoothing effect of the d$(V/V_{max})=0.1$ binning used for the deeper
\textit{r} band sample which may smooth over local variations in the density
profile. We therefore proceed to investigate \textit{r}-band LF's and the
resulting density profiles.

\begin{figure}
%
    \centering
\includegraphics[width=0.5\textwidth]{%
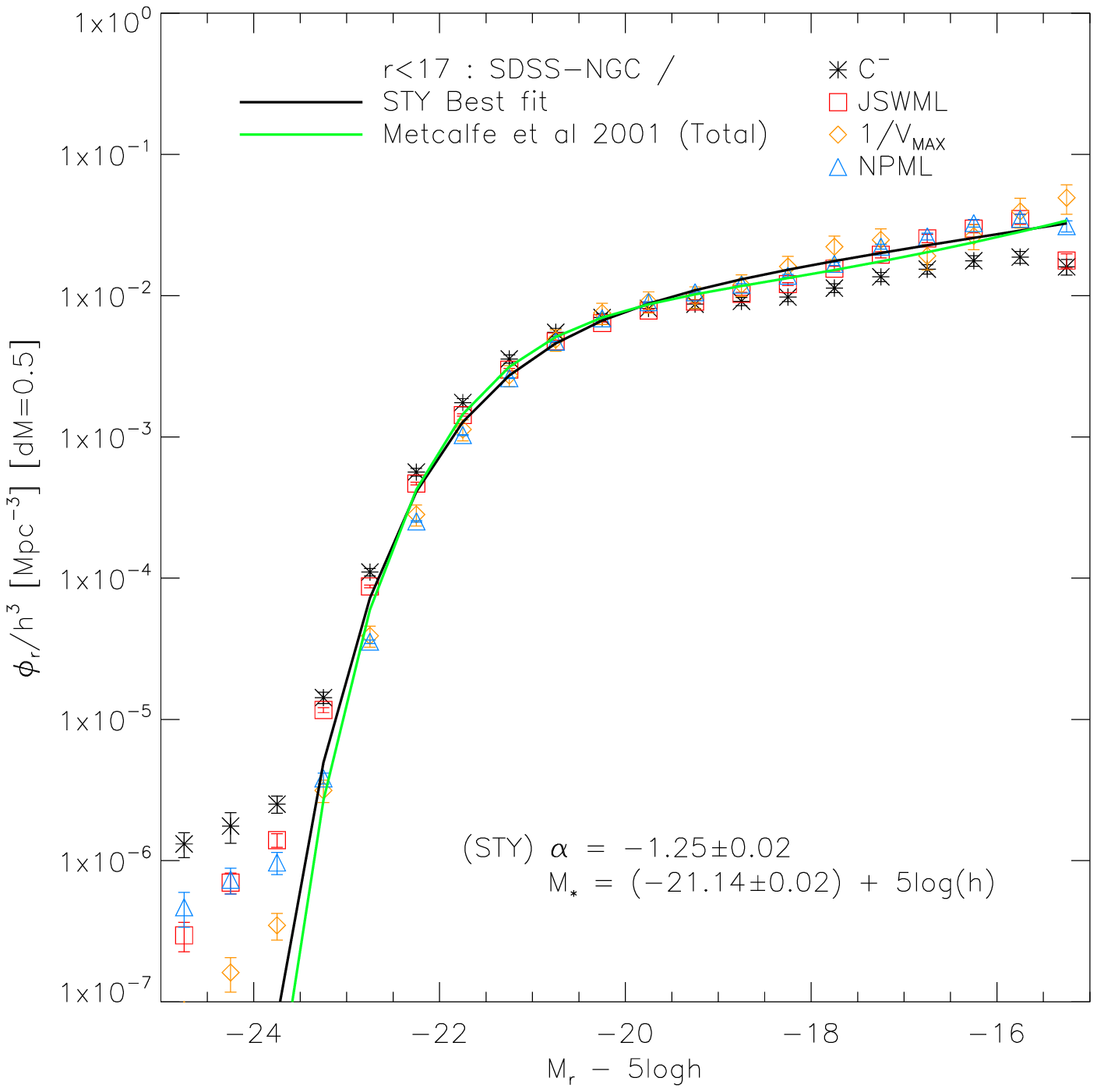}
    \caption[]{\textit{r} band galaxy LF estimates with $10<r<17$ and d$M = 0.5$
for the SDSS-NGC data. All of the estimates have been normalised using the
$n_{1}$ number density estimator for their respective LF.}
    \label{fig:lf_sdss_north_r}
\end{figure}

In Fig. \ref{fig:lf_sdss_north_r} we show the LF estimates for the SDSS-NGC region
\textit{r} band data. We see that there is good agreement between the variety of
LF estimates except for the very brightest objects ($<M_{*}-3$). We also note
that the $C^{-}$ estimator has a shallower faint end slope than compared to our
other LF estimators. However, again, these differences are relatively minor in
that the Metcalfe et al LF is a good representation of our results over a wide
range of magnitudes.

\begin{figure}
    \centering
\includegraphics[width=0.5\textwidth]{%
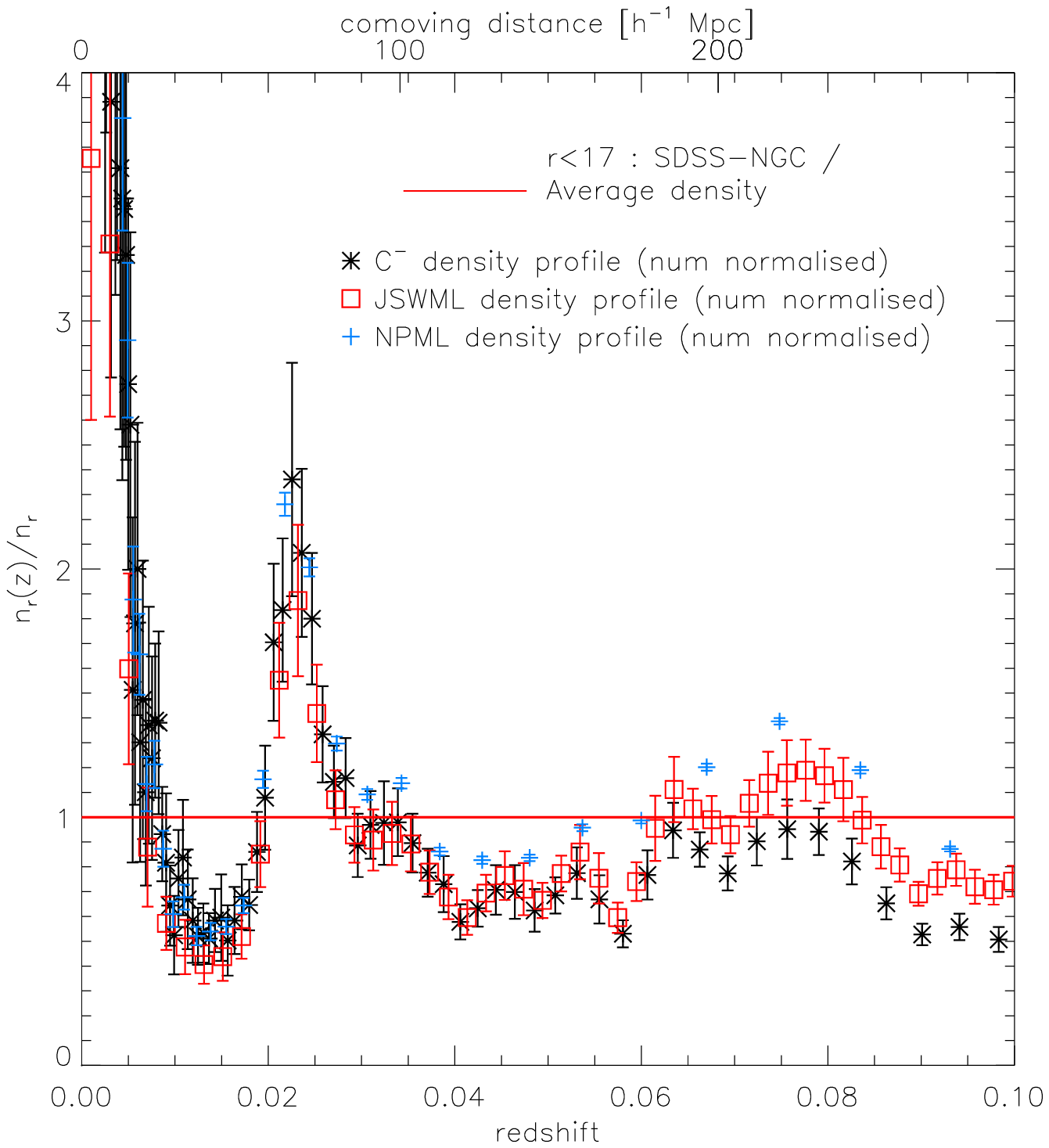}
    \caption[]{\textit{r} band galaxy number density profile with $10<r<17$ for
the SDSS-NGC data normalised (for each LF) to the number counts in the region.
The red (solid) line represents the homogeneous prediction for each LF.}
    \label{fig:nz_sdss_north_r4}
\end{figure}

We therefore show in Fig. \ref{fig:nz_sdss_north_r4} the density profiles
associated with JSWML, $C^{-}$ and NPML LF estimates \textit{r}-band SDSS-NGC
density profiles. The JSWML, $C^{-}$ and NPML profiles are in good agreement in
showing underdense profiles with significant LSS (which we attribute to Coma
at $z=0.023$). At deeper redshifts ($z>0.08$) the $C^{-}$, NPML and JSWML
density profiles remain significantly inhomogeneous. This is consistent with the
investigation of fainter GAMA \textit{K} band and SDSS \textit{r} band $n(m)$
and $n(z)$ in Paper I which indicated that an inhomogeneity could extend beyond
$z=0.1$ in the SDSS-NGC region. We also note that the \textit{r}-band density
profiles demonstrate a similar local underdensity ($z<0.08$) to that seen over
the same field in the \textit{K} band (see Fig. \ref{fig:nz_sdss_north_K3}).

Finally, we have evaluated the average number underdensity following
equation (\ref{eq:density_evaluate}) with the results shown in Table
\ref{tb:underdensity_table_r}. We now include the NPML density profiles for the
$z<0.1$ results as there is no evidence of lack of robustness at high redshift
for the \textit{r} band sample. For $z<0.05$ we find an average number
underdensity of $[0.83\pm0.05]$ which is in agreement with the \textit{K} band
SDSS-NGC results in suggesting a significant local number underdensity
($\approx150$h$^{-1}$Mpc). Whilst the $z<0.1$ average number underdensity of
$[0.90\pm0.03]$ indicates a more extensive inhomogeneity on
$\approx300$h$^{-1}$Mpc scales. Both these results are in agreement with the
results presented in Paper I.

We conclude that the \textit{r}-band density profiles show evidence for
an underdensity, which is punctuated by the Coma cluster producing a strong
overdensity. This underdensity is similar to those observed for the
corresponding region in the \textit{K} band (Fig. \ref{fig:nz_sdss_north_K3})
but less than that observed in the \textit{K} band over the SGC (Fig.
\ref{fig:nz_sixdf_south_K2}).

%
%

\begin{table}
 \begin{center}
 \begin{tabular}{ccc}

\hline
Field & Sample limit & Under-density \\
\hline

SDSS-NGC & $z<0.05$ & $0.83\pm0.05$ \\

\hline

SDSS-NGC & $z<0.10$ & $0.90\pm0.03$ \\

\hline
\end{tabular}
\caption{A summary of the average underdensities derived using eqn.
(\ref{eq:density_evaluate}). The $z<0.05$ and $z<0.1$ entries use
$10<r<17$.
 }
\label{tb:underdensity_table_r}
 \end{center}
\end{table}
\section{Discussion}
\label{sec:discussion_lf}

We note that any estimation of the LF is strongly dependent on
accurate and stable galaxy photometry. Paper I includes a fuller discussion of
the photometric completeness of these samples. However, we have estimated the
effect of magnitude errors on our NIR LF estimation using the method of
\citet{efstathiou_1988} where the observed LF is described as the
convolution of the true underlying LF with a magnitude error kernel, i.e.
$\phi_{obs}(M) = g(M) \otimes \phi_{true}(M)$. Using realistic magnitude error
kernels derived from fig. A1 of Paper I we find that the effects
of magnitude errors on the \textit{K}-band LF typically steepen the faint end
slope by $\approx0.1$ and similarly brighten the characteristic magnitude by
$\approx0.1$. We have investigated the uncertainty this corresponds to in our
determination of the galaxy number density profiles by testing the variation induced
in the \citet{cole_2011} density profiles if a realistic magnitude
error is allowed. We found that the changes were small, typically
$<1-2\%$ (\textit{K} and \textit{r} band respectively) and random in nature. It
should be noted that larger ($\approx 5\%$) variations were possible for low and
high z ($z<0.01$ and $z>0.1$). However, over the redshift range of interest we
conclude that magnitude errors only weakly affect our density profile estimates.

We again follow Paper I in the treatment of completeness issues.
We note that the inclusion of corrections for incompleteness detailed
for the $V_{max}$, $C^{-}$ and STY methods are relatively minor in determining
the LF or number density profiles. Finally, we also again note that we have
used the $vc$ 2MASS quality flag to reject artifact and non-extragalactic
objects. This ensures that $\approx$100\% of objects $7.5<K<12.5$ have been
visually inspected to ensure a high purity sample.

An important assumption in this work has been the use of the $K+E$
prescriptions used in \citet{metcalfe_2001,metcalfe_2006}. In order to explain
the observed underdensity we would require evolutionary brightening at
$z\approx0.1$ or a more negative K-correction. As was noted in Paper I an
argument against the rise in number density being caused by $z\approx0.1$ galaxy
evolution is the rise in number counts across the NIR and optical bands
(\textit{B},\textit{R},\textit{I},\textit{H},\textit{K})
\citet{metcalfe_2001,metcalfe_2006}. A local underdensity produces
just this, an approximately band-independent rise in the bright number counts
whereas evolutionary effects correspond to a greater effect in the bluer bands
and at fainter magnitudes. We also note that in Paper I we
investigated alternative evolutionary models as well as no evolutionary
corrections and found minimal differences in the \textit{K}-band (although not
for the \textit{r}-band) in terms of derived redshift distributions and number
counts.

The three fields studied here are wide field, with each representing $\approx
1/15$ of the sky. However, they are considerably smaller than the full sky 2MASS
sample from which they are drawn. We note however that \citet{appleby_2014} have
investigated the isotropy of LF shape estimates using the 2MPZ, a set of
photometric redshifts estimated in \citet{bilicki_2014} for the 2MASS-XSC
sample. These authors find no significant evidence for anisotropy in
non-parametric LF shape estimates. This suggests that the three fields used in
this study should be representative of the full 2MASS survey. It should be noted
that the LF normalisation was not investigated in this paper so this result is
not in tension with the varying galaxy number density profiles presented in this study.
Furthermore, it is also of interest that these authors report weak evidence for
a dipole asymmetry in parametric LF estimates between the north and south
galactic plane. This is again in agreement with the significantly different
density profiles found for the 6DFGS-SGC, 6DFGS-NGC and SDSS-NGC regions.

\begin{figure}
   \begin{minipage}{0.5\textwidth}
   \centering
\includegraphics[width=1.0\textwidth]{%
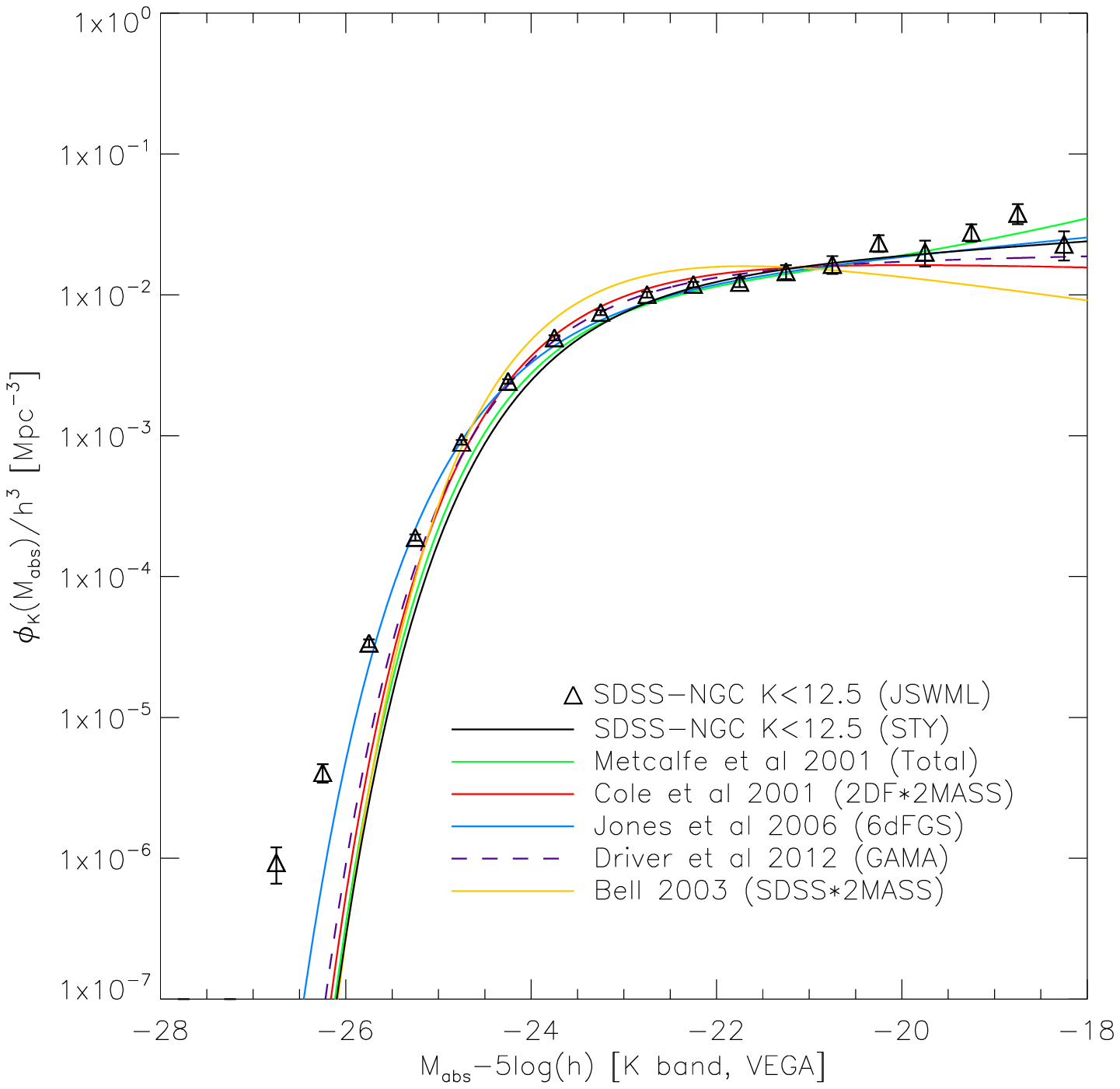}
 \caption[A comparison of various \textit{K} band LF estimates using the full
$\phi_{K}(M)$.]{A comparison of various parametric \textit{K} band LF estimates
using the full $\phi_{K}(M)$. Results have been normalised to a common and
arbitrary number density estimated using the Metcalfe et al LF over the range
$-28\le M-5\log(h) \le -18$. Parameters taken from \citet{cole_2001},
\citet{jones_2006}, \citet{driver_2012} and \citet{bell_2003}.
}
 \label{fig:lf_fullphi_k}
    \end{minipage}
    \hspace{1pc}%
    \hfill
    \\ \\ \\
    \begin{minipage}{0.5\textwidth}
    \centering
\includegraphics[width=1.0\textwidth]{%
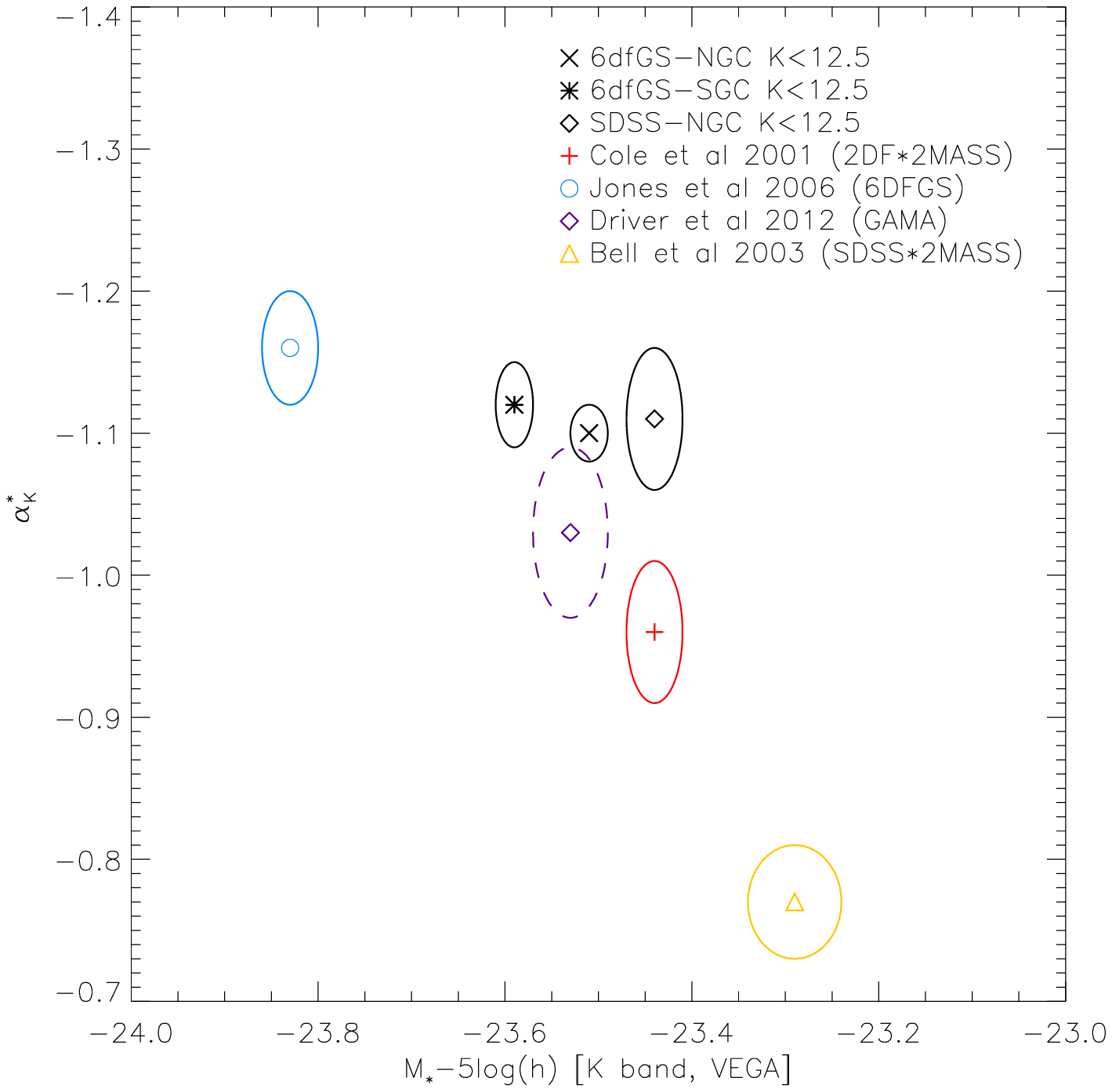}
 \caption{A comparison of various \textit{K} band parametric LF estimates of the
Schechter $\alpha$ and $M_{*}$ parameters. We have not been able to represent
the covariance between $\alpha$ and $M_{*}$ from the studies shown so we assume
no covariance.
}
 \label{fig:lf_alphaMstar_k}
\end{minipage}
\end{figure}

Other estimates of the NIR LF using these samples have been attempted. We
therefore now present in Figs \ref{fig:lf_fullphi_k} \&
\ref{fig:lf_alphaMstar_k} a comparison to other studies estimates of the
Schechter $\alpha$ and $M_{*}$ parameters. In Fig. \ref{fig:lf_fullphi_k} we
show a comparison of the full $\phi(M)$ because it captures the correlation
between the $\alpha$ and $M_{*}$ parameters. We also include in this comparison
our JSWML non parametric LF estimates (SDSS-NGC) so that any deviations from the
Schechter form can be judged. After normalising to a common and arbitrary
number density [estimated using the Metcalfe et al LF over the range $-28\le
M-5\log(h) \le -18$] we find that the \textit{K}-band LF estimates are
relatively consistent, except at the faint end where there is greater variance.
In particular, the Metcalfe et al LF shows a much steeper faint-end slope than
found by \citet{bell_2003} using the $V_{max}$ estimator with the 2MASS survey.
We attribute this to photometric problems in the early 2MASS data releases since
in the same work the \textit{g}-band selected NIR LF was estimated to have a
faint end slope of $\alpha=-1.33$ which is in rough agreement with the steeper
slope of the Metcalfe et al LF. 

In Fig. \ref{fig:lf_alphaMstar_k} most \textit{K}-band LF  parameterisations 
occupy the usual degenerate strip between $M^*$ and $\alpha$. The one that most
deviates from this line is the LF of \citet{jones_2006} (see also
Fig.\ref{fig:lf_fullphi_k}). We also note that the variation between different
LF estimates is much larger than would be expected on the basis of the `naive'
error ellipses (i.e. the no covariance case). We do not understand what causes
these differences but potential causes of differences are different treatment of
flow models, incompleteness, as well as real differences in sample selection
(magnitude limits, redshift ranges, etc.).

\begin{figure}
\begin{minipage}{0.5\textwidth}
   \centering
\includegraphics[width=1.0\textwidth]{%
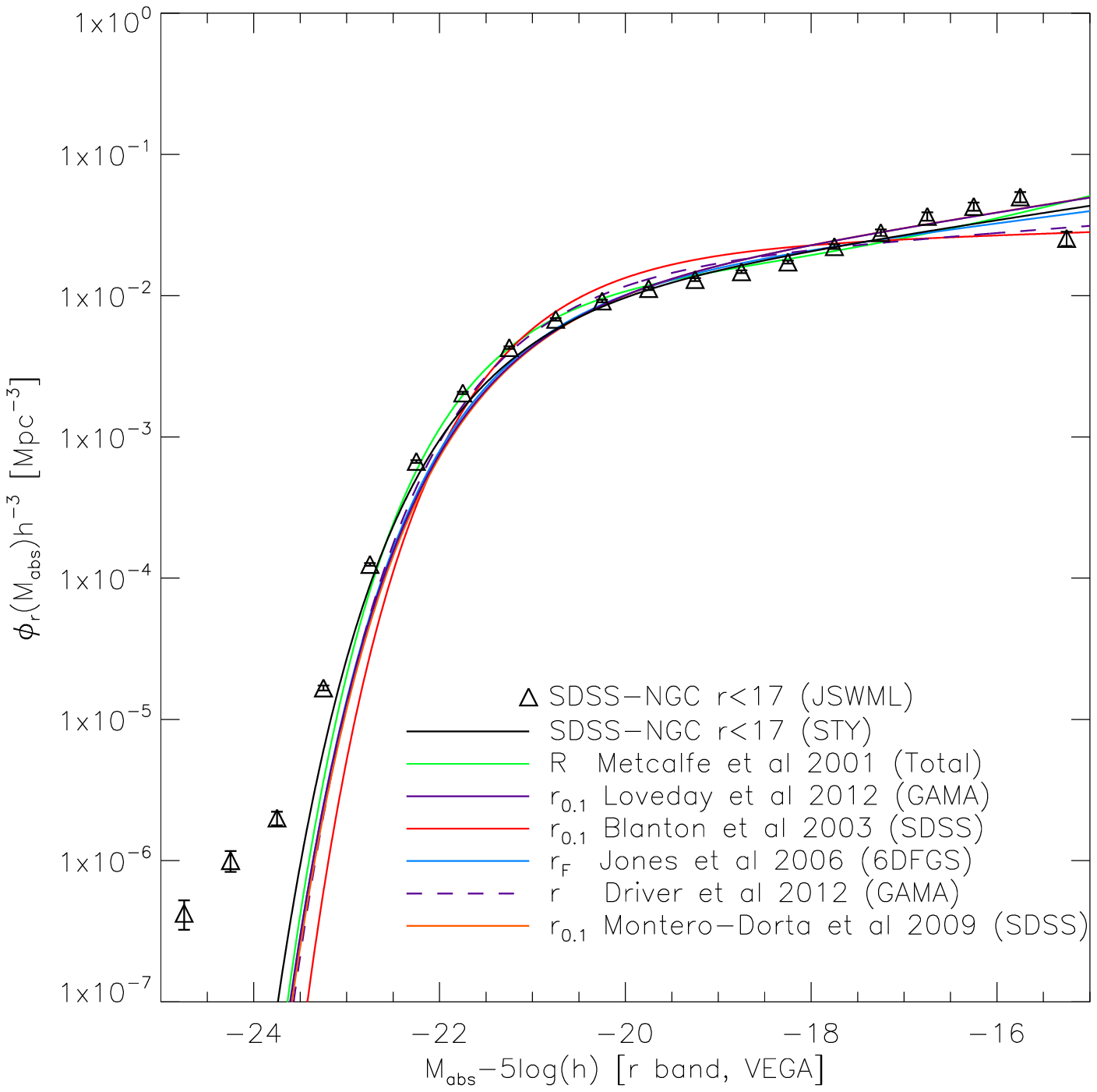}
 \caption{A comparison of various \textit{r} band LF estimates using the full
$\phi_{r}(M)$. Results have been normalised to a common and arbitrary
number density estimated using the Metcalfe et al LF over the range $-25\le
M-5\log(h) \le -15$. Parameters taken from \citet{loveday_2012},
\citet{jones_2006}, \citet{blanton_2003}, \citet{driver_2012} and
\citet{montero_2009}. Results quoted in the $r_{0.1}$ band have been converted
into the $r$ band as $r_{0.1} \approx r+0.23$ \citep{nichol_2006}.
}
 \label{fig:lf_fullphi_r}
    \end{minipage}
    \hspace{1pc}%
    \hfill
    \begin{minipage}{0.5\textwidth}
    \centering
\includegraphics[width=1.0\textwidth]{%
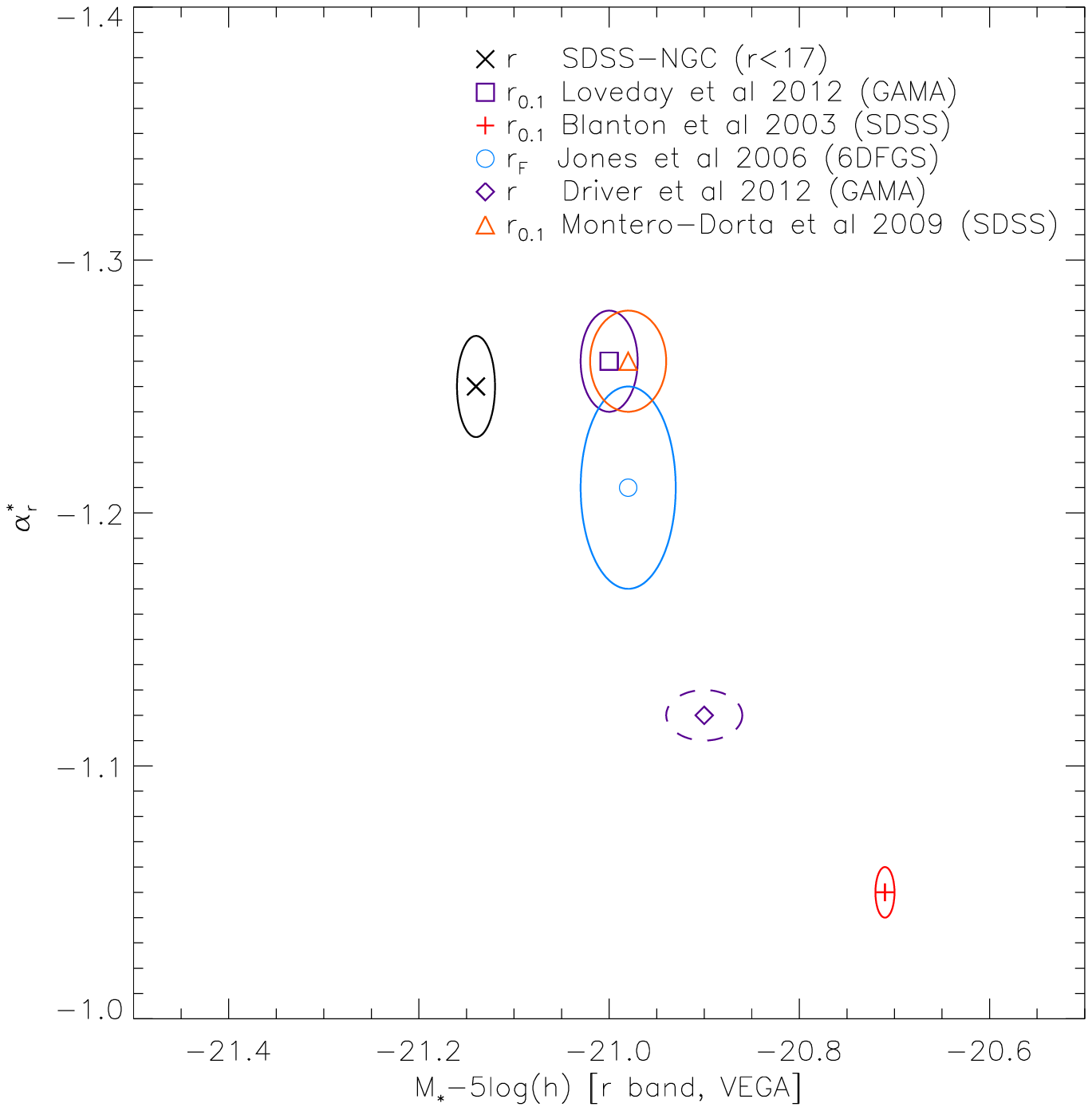}
 \caption[A comparison of various \textit{r} band parametric LF estimates of the
Schechter $\alpha$ and $M_{*}$ parameters.]{A comparison of various r band
parametric LF estimates of the Schechter $\alpha$ and $M_{*}$ parameters.
We have not been able to represent the covariance between $\alpha$ and
$M_{*}$ from the studies shown so we assume no covariance. Again, results quoted
in the $r_{0.1}$ band have been converted into the $r$ band as $r_{0.1}
\approx r+0.23$
\citep{nichol_2006}
}
 \label{fig:lf_alphaMstar_r}
\end{minipage}
\end{figure}

We have made a similar comparison for \textit{r} band LF estimates in Figs
\ref{fig:lf_fullphi_r} \& \ref{fig:lf_alphaMstar_r}. We have again
normalised to an arbitrary number density estimated using the Metcalfe
et al LF over the range $-25\le M-5\log(h) \le -15$. Results quoted in
the $r_{0.1}$ band (i.e. corrected to a $z=0.1$ rest frame as per
\citealt{blanton_2003}) have been converted into the $r$ band as $r_{0.1}
\approx r+0.23$ \citep{nichol_2006}. We find results that are consistent
with the Metcalfe et al LF. However, the \citet{blanton_2003} and
\citet{driver_2012} estimates show a significantly sharper bright end
fall off and a shallower faint end slope than the Metcalfe et al LF. The
greater uncertainties in the assumed evolutionary model for the \textit{r}-band
as compared to the \textit{K} may explain the difference with the
\citet{driver_2012} LF since the GAMA survey probes a substantially
deeper redshift range than the samples used in this paper. However, this
is an unlikely explanation for the difference with the SDSS based
\citet{blanton_2003} LF, especially as any evolutionary modelling
effects would be expected to primarily affect $M_{*}$ estimates. It is
therefore unclear why these results are different from those presented
here. However, we agree with the observation of \citet{montero_2009}
that the size of the SDSS sample has increased considerably since the
pre SDSS-DR1 results used in \citet{blanton_2003}. The resulting
improvements in magnitude limits may have resulted in substantial
improvements in sample selection. The potential uncertainty that differences in
the LF correspond to in the number density profiles is indicated by the
differences between the $C^{-}$, JSWML and NPML density profiles in Fig.
\ref{fig:nz_sdss_north_r4}. This is particularly relevant in the case of the
$C^{-}$ LF estimate where we find a flatter faint end slope than the JSWML or
NPML estimates (see Fig. \ref{fig:lf_sdss_north_r}) but nevertheless find
similar number density profiles.


Again many, r-band  LF  parameterisations occupy the usual degenerate
strip between $M^*$ and $\alpha$ in Fig. \ref{fig:lf_alphaMstar_r}. The
two that most deviate from this line are the LF's  of
\citet{loveday_2012} and \citet{montero_2009} (see also 
Fig.\ref{fig:lf_fullphi_r}). However, this small deviation may be
attributable to the correction applied to convert from the $r_{0.1}$ and
\textit{r} bands. Fig. \ref{fig:lf_alphaMstar_r} also confirms the much flatter
parametric faint end slope reported by \citet{blanton_2003} than seen by later authors, including ourselves.

\section{Conclusions}
\label{sec:conclusion_lf}

We have used samples from 6dFGS/2MASS and SDSS to simultaneously
investigate the local LF and galaxy number density profiles. We have
studied three large volumes which cover much of the sky and find
evidence for an anisotropic galaxy distribution. In particular we
observe a local number underdensity out to $\approx150$h$^{-1}$Mpc
around our position in the SGC which both the \textit{r} and \textit{K}
band SDSS-NGC samples suggest may extend deeper to $\approx300$h$^{-1}$Mpc.
We have also found evidence that the Metcalfe et al LF assumed in Paper
I is an adequate fit for these samples and hence the density profiles
presented there may be unbiased by this choice. This work also
complements previous studies which have investigated variation of
luminosity density with redshift \citep{keenan_2012,keenan_2013} by
providing estimates of variation in number density. The estimate made in
Paper I of an $\approx 15\%$ number underdensity is broadly consistent
with the \citet{keenan_2013} estimate of an $\approx 50\%$ increase in
luminosity between the local universe and $z\approx0.1$.

A significant advantage of investigating both the \textit{K} and \textit{r}
bands is that an underdensity might be indicated if the effect is band
independent, although a band-dependent result might still be explained by
different galaxy clustering bias applying in the different bands. We note that
the \textit{r}-band SDSS-NGC profile (Fig. \ref{fig:nz_sdss_north_r4}) shows a
similar underdensity to the corresponding \textit{K}-band estimate (Fig.
\ref{fig:nz_sdss_north_K3}) with only small differences at low redshift. But
both are in agreement in supporting an underdensity continuing beyond $z \approx
0.1$. One important route for continuing the investigation into the
isotropy of the galaxy distribution will be the incorporation of 
peculiar velocity fields. We therefore believe that the ongoing
investigation into the 6DFGS peculiar velocity field as determined using
the Fundamental Plane (and its comparison to that inferred from PSCz)
started in \citet{springob_2014} will be of particular interest. We note
that the initial analysis presented in \citet{springob_2014} is in
agreement with the PSCz estimated density field (to a separately
estimated 2MRS peculiar velocity field based on an update of
\citealt{erdogdu_2006}) which has a mean underdensity of $11\%$ within
$180h^{-1}$Mpc. The density profiles presented here are consistent with
such a local underdensity.


\section*{Acknowledegments}
\label{sec:acknowledgements}

We thank Richard Fong, John Lucey, Alan Heavens and Ryan Keenan for useful
comments. JRW acknowledges financial support from STFC. We would also like to
thank the anonymous referee for their comments.

We acknowledge the use of a modified version of the \texttt{JSWML} code
accompanying \citet{cole_2011}. 

This research has made use of the NASA/IPAC Extragalactic Database (NED) which
is operated by the Jet Propulsion Laboratory, California Institute of
Technology, under contract with the National Aeronautics and Space
Administration.

We would also like to acknowledge the use of the \texttt{TOPCAT} utility
\citep{2005ASPC..347...29T}. 

This research has made use of the VizieR catalogue access tool, CDS, Strasbourg,
France''. The original description of the VizieR service was published in A\&AS
143, 23 (2000).

Funding for the SDSS and SDSS-II has been provided by the Alfred P. Sloan
Foundation, the Participating Institutions, the National Science Foundation, the
U.S. Department of Energy, the National Aeronautics and Space Administration,
the Japanese Monbukagakusho, the Max Planck Society, and the Higher Education
Funding Council for England. The SDSS Web Site is http://www.sdss.org/.

The SDSS is managed by the Astrophysical Research Consortium for the
Participating Institutions. The Participating Institutions are the American
Museum of Natural History, Astrophysical Institute Potsdam, University of Basel,
University of Cambridge, Case Western Reserve University, University of Chicago,
Drexel University, Fermilab, the Institute for Advanced Study, the Japan
Participation Group, Johns Hopkins University, the Joint Institute for Nuclear
Astrophysics, the Kavli Institute for Particle Astrophysics and Cosmology, the
Korean Scientist Group, the Chinese Academy of Sciences (LAMOST), Los Alamos
National Laboratory, the Max-Planck-Institute for Astronomy (MPIA), the
Max-Planck-Institute for Astrophysics (MPA), New Mexico State University, Ohio
State University, University of Pittsburgh, University of Portsmouth, Princeton
University, the United States Naval Observatory, and the University of
Washington.

This publication makes use of data products from the Two Micron All Sky Survey,
which is a joint project of the University of Massachusetts and the Infrared
Processing and Analysis Center/California Institute of Technology, funded by the
National Aeronautics and Space Administration and the National Science
Foundation.

\setlength{\bibhang}{2.0em}
\setlength\labelwidth{0.0em}
\bibliographystyle{mn2e_alt}
\bibliography{%
/obs/r1/Dropbox/astro_writing/thesis_draft/thesis_draft_3/bibtex_localcounts,%
/obs/r1/Dropbox/astro_writing/thesis_draft/thesis_draft_3/bibtex_planckwmap}

\bsp
\label{lastpage}
\end{document}